\documentclass[runningheads]{llncs}

\usepackage[margin=1in]{geometry}
\usepackage[T1]{fontenc}
\usepackage{graphicx}
\usepackage{color}
\usepackage{amssymb}
\usepackage{fancyhdr}
\usepackage{placeins}
\usepackage[linesnumbered,ruled,vlined]{algorithm2e}

\SetCommentSty{mycommfont}
\SetKwInput{KwInput}{Input}                
\SetKwInput{KwOutput}{Output}              
\SetKwInput{KwFunctions}{Functions}         
\SetKwInput{KwTrainables}{Trainable Parameters}
\SetKwInput{KwStopgrad}{Stop Gradient} 
\SetKwInput{KwHyperparameters}{Hyperparameters}    
\SetKwInput{KwSamples}{Samples}  

\usepackage{tablefootnote}
\usepackage{booktabs, multirow} 
\usepackage{soul}
\usepackage[table]{xcolor} 
\usepackage{changepage,threeparttable} 
\usepackage[colorlinks=true,urlcolor=blue,linkcolor=blue,citecolor=blue]{hyperref}
\usepackage{array}
\usepackage{xcolor}
\usepackage{makecell}
\usepackage{epigraph}
\usepackage[style=ieee]{biblatex}
\usepackage{adjustbox}
\def\figurename{Fig.}

\def\tablename{Table}
\newcolumntype{P}[1]{>{\centering\arraybackslash}p{#1}}
\definecolor{maroon}{RGB}{58, 183, 149}
\definecolor{iblue}{RGB}{33, 53, 236 }
\definecolor{tablered}{cmyk}{0,0.87,0.68,0.32}

\begin{document}
\title{BrainSegFounder: Towards Foundation Models for Neuroimage Segmentation}
\titlerunning{BrainSegFounder: Towards Brain Segmentation Foundation Models}

\author{Joseph Cox\inst{1} \and 
 Peng Liu\inst{1} \and
 Skylar E. Stolte\inst{1} \and
 Yunchao Yang\inst{1} \and
 Kang Liu\inst{1} \and
 Kyle B. See\inst{1} \and
 Huiwen Ju\inst{2} \and
 Ruogu Fang\inst{1}$^{,*}$
 } 
\authorrunning{J. Cox et al. }
\institute{University of Florida, Gainesville, FL 32611, USA \\
 \email{\{cox.j,plui1,skylastolte444,yunchaoyang,\\kang.lui,kylebsee,ruogu.fang\}@ufl.edu} \and
 NVIDIA, Santa Clara, CA 95051, USA \\
\email{hju@nvidia.com}}

\maketitle              

\thispagestyle{fancy}
\fancyfoot[L]{$^*$ Corresponding Author}
\begin{abstract}
The burgeoning field of brain health research increasingly leverages artificial intelligence (AI) to analyze and interpret neuroimaging data. Medical foundation models have shown promise of superior performance with better sample efficiency. This work introduces a novel approach towards creating 3-dimensional (3D) medical foundation models for multimodal neuroimage segmentation through self-supervised training. Our approach involves a novel two-stage pretraining approach using vision transformers. The first stage encodes anatomical structures in generally healthy brains from the large-scale unlabeled neuroimage dataset of multimodal brain magnetic resonance imaging (MRI) images from 41,400 participants. This stage of pertaining focuses on identifying key features such as shapes and sizes of different brain structures. The second pretraining stage identifies disease-specific attributes, such as geometric shapes of tumors and lesions and spatial placements within the brain. This dual-phase methodology significantly reduces the extensive data requirements usually necessary for AI model training in neuroimage segmentation with the flexibility to adapt to various imaging modalities. We rigorously evaluate our model, BrainSegFounder, using the Brain Tumor Segmentation (BraTS) challenge and Anatomical Tracings of Lesions After Stroke v2.0 (ATLAS v2.0) datasets. BrainSegFounder demonstrates a significant performance gain, surpassing the achievements of the previous winning solutions using fully supervised learning. Our findings underscore the impact of scaling up both the model complexity and the volume of unlabeled training data derived from generally healthy brains. Both of these factors enhance the accuracy and predictive capabilities of the model in neuroimage segmentation tasks. Our pretrained models and code are at \url{https://github.com/lab-smile/BrainSegFounder}.
\end{abstract}

\keywords{Neuroimaging \and 3D Foundation Model \and Self-Supervised Learning \and Brain Tumor Segmentation \and Multi-modal MRI}

\section{Introduction}

The fusion of artificial intelligence (AI) with neuroimaging analysis, particularly multimodal MRI, is forging a pivotal role in advancing brain health (\cite{chen_ai-based_2022}, \cite{segato_artificial_2020}, \cite{rao_brain_2023}, \cite{owolabi_global_2023}, \cite{moreno-blanco_technologies_2019}, \cite{rajpurkar_ai_2022} \cite{khachaturian_accelerating_2023}). The complexity of the human brain, with its elaborate anatomy and intricate functions, poses significant challenges in neuroimaging analysis (\cite{moor_foundation_2023}, \cite{azad_foundational_2023},\cite{zhang_challenges_2024}, \cite{segato_artificial_2020}, \cite{rajpurkar_ai_2022}). AI's capability to interpret complex neurological data has the potential to enhance diagnostic precision and deepen our understanding of brain pathology. Numerous studies have aimed to develop AI models for specific brain health analyses, each contributing to the growing body of neuroimaging research.
 
Traditionally, neuroimaging AI models require extensive fine-tuning through supervised learning to address a specific downstream task. Modifications of the nnU-Net (\cite{isensee_nnu-net_2021}), DeepScan (\cite{mckinley_ensembles_2019}), and DeepMedic (\cite{kamnitsas_efficient_2017}) architectures have performed well on a host of medical computer vision challenges such as the Brain Tumor Segmentation (BraTS) challenge (\cite{baid_rsna-asnr-miccai_2021}), Medical Segmentation Decathlon (MSD) (\cite{antonelli_medical_2022}), and A tumor and liver automatic segmentation challenge (ATLAS) (\cite{Quinton2023}). Many of these advances stem from utilizing self-supervised pretraining methods on large, unlabeled datasets to transfer weights for model encoders and decoders to the smaller datasets present in the challenge (\cite{zhou_models_2021}, \cite{tang_self-supervised_2022}). Complementary to these pretraining modifications, there has been a recent push towards developing massive medical datasets (\cite{mei_radimagenet_2022}, \cite{clark_cancer_2013}, \cite{bycroft_uk_2018}) to aid in the creation of these models. However, medical image analysis has yet to benefit from the recent advances in natural image analysis and language processing through models like the Segment Anything Model (SAM) (\cite{kirillov_segment_2023}) and LLaMA (\cite{touvron_llama_2023}).  
 
In medical language processing, models like MI-Zero (\cite{lu_visual_2023}) and BioViL-T (\cite{bannur_learning_2023}) utilize contrastive learning to make significant advancements in representational analysis and zero-shot transfer learning in medical image recognition. By leveraging different learning objectives, similar image-text pairs are pulled closer in the latent space while dissimilar pairs are pushed further apart. Such models have pushed the boundary of histopathology research and combined text-based analysis with computer vision. Yet, they rely on having text-based prompts accompanying their training images (\cite{tiu_expert-level_2022}).  

With SAM’s demonstrated success on few-shot segmentation tasks of natural images, recent research into medical image segmentation models has primarily modified the SAM architecture. Models like MedSAM (\cite{ma_segment_2023}), MedLSAM (\cite{lei_medlsam_2023}), and SAM-Med2D (\cite{cheng_sam-med2d_2023}) focus on bridging the gap between SAM’s generalizability on real-world images and its performance on medical tasks. They accomplish this by adapting the SAM architecture to these medical tasks. \cite{ma_segment_2023} crafted MedSAM for image segmentation by constructing a massive dataset of image-mask pairs derived from sizeable medical image databases. MedLSAM further refined upon MedSAM by including landmark localization. SAM-Med2D further improved segmentation results by increasing the dataset to multiple modalities and increasing prompt density. However, these models function in 2-dimensional space, requiring 3-dimensional (3D) modalities to be sub-sampled or solved in slices (\cite{azad_foundational_2023}). Not only is this computationally inefficient, but the most dense and often most valuable information is found in 3D modalities like CT or MRI. \cite{Gong2023} aimed to address this discrepancy by adapting the SAM models to 3D space using a visual sampler and a mask decoder to aggregate layers. Their model, dubbed 3DSAM-adapter, outperformed leading segmentation models in various tasks while still utilizing an algorithm that functions in 2D space. These results indicated that these models would benefit from the critical anatomical and spatial information found from being fully capable of functioning in 3D space.

Despite the progress in medical imaging, holistically analyzing the vast amount of data generated by brain MRIs remains a formidable challenge (\cite{azad_foundational_2023}). The intricate structure and function of the brain necessitate advancements in MRI analysis due to their critical impact on patient outcomes, especially in the early detection and treatment of brain disorders (\cite{zhang_challenges_2024}). Existing AI models in neuroimaging are hampered by their need for extensive supervised learning and their limited ability to generalize across different tasks without substantial retraining, revealing a gap for a robust, adaptable model that functions in 3D space (\cite{azad_foundational_2023}, \cite{zhang_challenges_2024}).
 
This study presents BrainSegFounder, a 3D foundational framework for multimodal neuroimage segmentation. BrainSegFounder is designed to pave the way towards setting new standards for the accuracy and efficiency of medical AI models. We focus our study on two essential tasks - brain tumor segmentation and brain lesion segmentation. A primary obstacle in creating AI models for brain tumor and brain lesion analysis is the scarcity of brain tumors within the general population. This scarcity significantly hampers the compilation of large-scale diseased patient datasets, which are essential for the supervised training of AI models. In response, the development process of BrainSegFounder incorporates a multi-stage approach to feature learning, specifically engineered to mitigate the challenges posed by data scarcity.
 
In its initial phase, BrainSegFounder leverages an extensive dataset from brain scans of 41,400 participants from the United Kingdom. This foundational step enables the framework to effectively encode generally healthy brain tissue structures, creating a detailed baseline of anatomical features from a predominantly healthy population. Subsequently, the framework's training shifts focus towards identifying disease-specific attributes, such as geometric shapes of tumors and lesions and spatial placements within the brain. This dual-phase methodology significantly diminishes the extensive data requirements usually necessary for AI model training in tumor detection. Moreover, it naturally expands the dataset available for the AI to learn from efficiently and straightforwardly, sidestepping the need for generating synthetic images. This approach mirrors the analytical techniques used by radiologists and has undergone thorough validation against the BraTS challenge and ATLAS 2.0 datasets, showcasing significant improvements over current models.
 
BrainSegFounder, our novel framework, represents a pivotal advance in neuroimaging analysis by laying the groundwork for a future of comprehensive foundation models in this field. BrainSegFounder is designed to be adaptable for various neurological tasks, including brain tumor segmentation, stroke localization, brain region segmentation, and the diagnosis of Alzheimer's disease. By utilizing a large dataset of brain imaging from a generally healthy population, BrainSegFounder sets the stage for transforming clinical workflows, aiming to enhance the speed and accuracy of diagnoses across a spectrum of neurological conditions. The contribution of this work is twofold: 1) BrainSegFounder leverages a large-scale multi-modal 3D neuroimaging dataset of generally healthy brain images to create a latent-space representation of healthy brain MRI; 2) it gains the ability to detect anomalies by training on anomaly-specific datasets with adaptability between imaging modalities. This approach effectively addresses the challenge of limited patient data in diseased neuroimaging analysis, establishing BrainSegFounder as a versatile framework in medical diagnostics. Its broad applicability signifies a shift towards more integrated and adaptable methodologies in neurological diagnostics.
 
\begin{figure*}[!t]
\centering
\includegraphics[width=\textwidth]{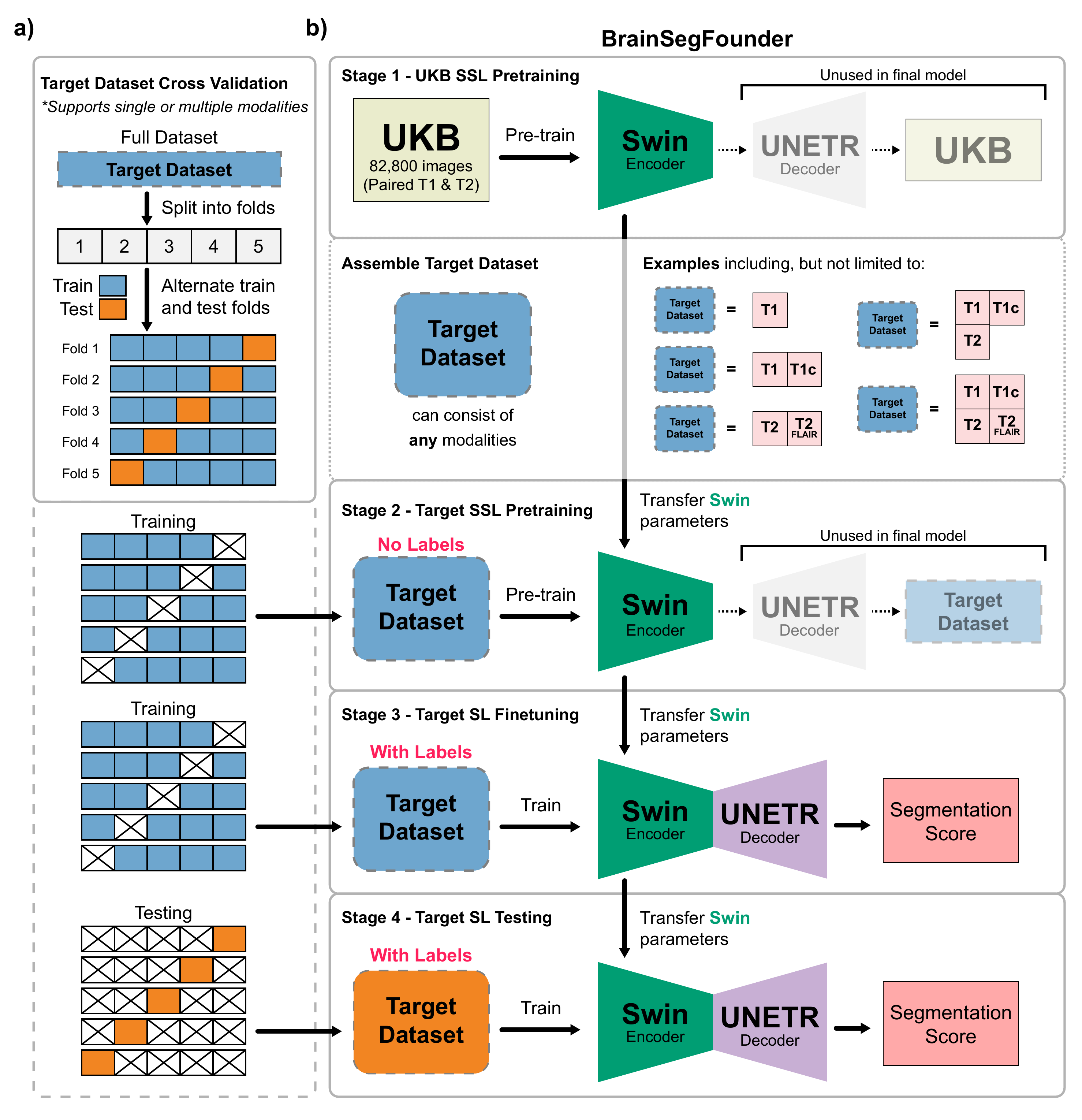}
\caption{Overall Study Design. a)  The two-stage pretraining process using Swin Transformer decoders and encoder. Initially, the model is pretrained on the UKB dataset (Stage 1), followed by the downstream task dataset (Stage 2). b) This is succeeded by fine-tuning on each downstream dataset, with transfer learning applied between each stage.}
\label{fig:overview}
\end{figure*}

 The BrainSegFounder framework introduces a deep learning training scheme tailored for diverse applications by showcasing a distinct approach to self-supervised pretraining followed by precise fine-tuning. This section offers a detailed examination of the framework's architecture and its procedural pipeline. It highlights the multi-stage self-supervised pretraining, termed Stage 1 and Stage 2, before proceeding to fine-tuning for downstream tasks. 
 Figure ~\ref{fig:overview} illustrates BrainSegFounder's architecture. Central to BrainSegFounder is a vision transformer-based encoder that employs a series of self-attention mechanisms. This encoder is linked with an up-sampling decoder tailored for segmentation tasks. The architecture is adapted from the SwinUNETR architecture \cite{hatamizadeh_swin_2022} with modified input channels and input hyperparameters. BrainSegFounder pioneers a novel dual-phase self-supervised pretraining method, integrating self-supervised learning components within its structure. Stage 1 pretraining exposes the framework to a wide-ranging dataset of brain MRIs from the UK Biobank dataset, predominantly consisting of healthy individuals. This initial stage equips the model with a thorough comprehension of standard brain anatomy, utilizing self-supervised learning to enhance prediction capabilities. Stage 2 of pretraining advances the model's proficiency by introducing it to a specialized MRI dataset geared toward the downstream task. This phase leverages the architecture's refined anomaly detection skills, focusing on distinguishing deviations in brain structure.

Following pretraining, BrainSegFounder undergoes fine-tuning on the final dataset, where transfer learning enhances the model's encoder. As depicted in Figure~\ref{fig:overview}, the fine-tuning process leverages the pretrained Swin Transformer encoder from the earlier two stages. The first pretraining stage on the UKB dataset develops a foundational understanding of normal brain anatomy. The second stage of pretraining with diseased datasets builds upon this foundation by introducing pathology, thus allowing the model to learn the distinction between healthy and pathological tissues. Transfer learning is applied after each pretraining stage to retain and refine the knowledge acquired, ensuring that the model can effectively adapt to the new dataset while preserving previously learned patterns.

 The culmination of this process is the integration of the U-NET decoder, which works in concert with the pretrained encoders to generate segmentation scores that delineate tumor boundaries with precision. This hybrid approach combines the strengths of the Swin Transformer and UNETR architectures, optimizing the model for the critical task of tumor segmentation and providing an authoritative score that reflects the model's accuracy in identifying and delineating tumor regions.

 In summary, the BrainSegFounder model's architecture and pretraining paradigm represent a comprehensive approach to understanding and segmenting brain images, with a training pipeline that methodically builds the model's capacity to differentiate and characterize complex patterns in 3D MRI data without external annotation. Together, the self-supervised learning stages and the fine-tuning process prepare BrainSegFounder to tackle downstream tasks with high efficiency and accuracy.

\subsection{Data Acquisition and Preprocessing}
 Throughout our pretraining and fine-tuning, we make use of the UK Biobank (UKB),  Brain Tumor Segmentation (BraTS) Challenge, and Anatomical Tracings of Lesions After Stroke v2.0 (ATLAS v2.0)  datasets. The following section summarizes the dataset information that is pertinent to our study. Table ~\ref{tab:datasets} provides an overview of this information.

\begin{table*}[t!]
    \centering
    \small 
    \setlength{\tabcolsep}{5pt}
    \renewcommand{\arraystretch}{1.2}

    \begin{tabular}{|l|l|l|l|}
        \hline
        & \textbf{UK Biobank}  & \textbf{BraTS} & \textbf{ATLAS} \\
        \hline
        Number of Subjects  & 41,400 & 1,251 & 655  \\
        \hline
        Modalities  & T1w, T2-FLAIR & T1w, T1-ce, T2w, T2-FLAIR & T1-ce \\
        \hline
        Number of Images & 82,800 & 5,004 & 655 \\
        \hline
        Diseases  & Generally Healthy & Malignant Brain Neoplasms & Stroke \\
        \hline
    \end{tabular}
\caption{A summary of the data used in this study.}
\label{tab:datasets}
\end{table*}

\subsubsection{UK Biobank dataset}

\begin{figure}[!t]
\centering
\includegraphics[width=0.5\columnwidth]{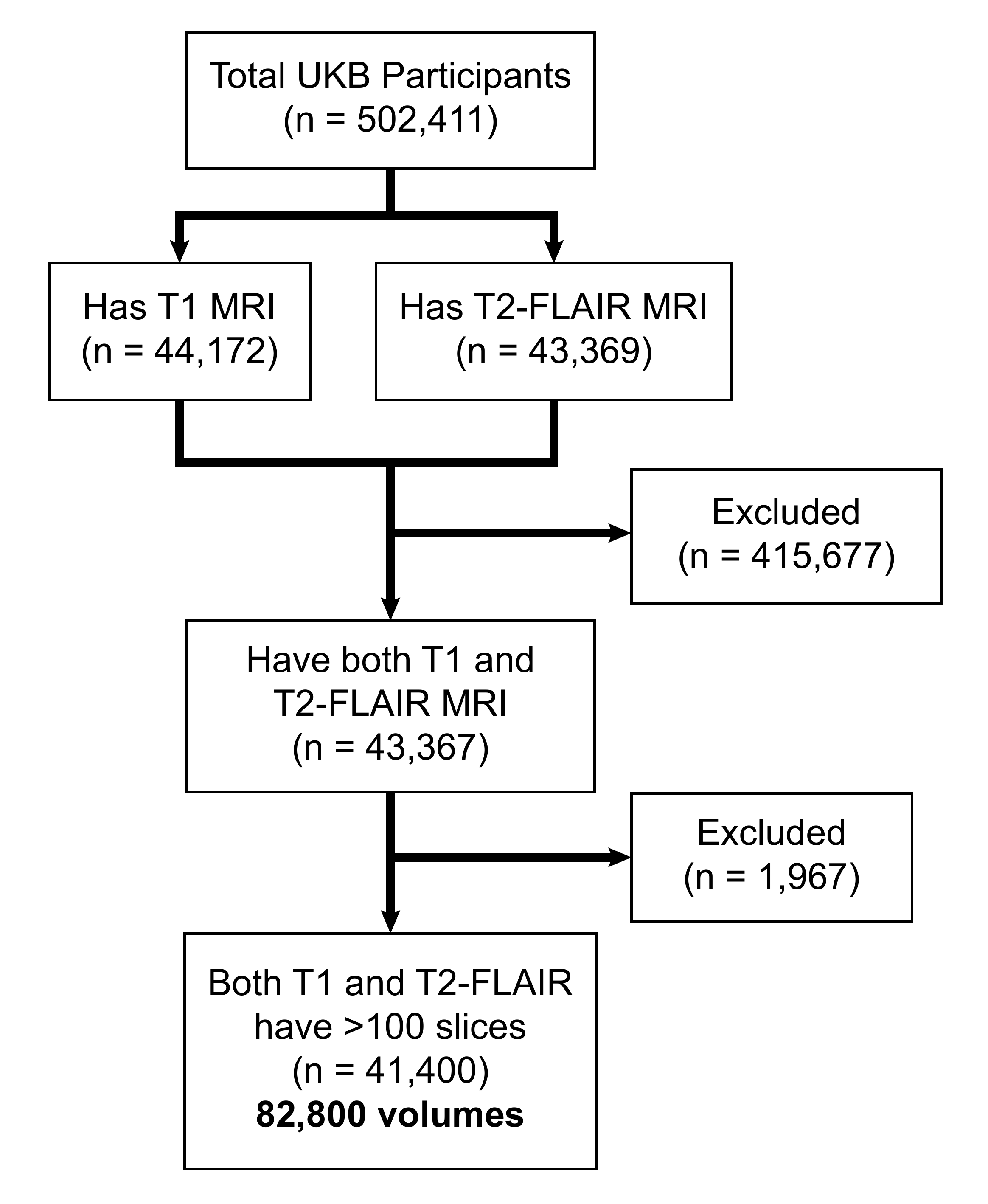}
\caption{CONSORT diagram of UKB data used in Stage 1 pretraining.}
\label{fig:consort}
\end{figure}

 In our first stage, we utilize T1-weighted (T1w) and T2-weighted Fluid Attenuation Inversion Recovery (T2-FLAIR) from the UK Biobank (UKB) dataset (\cite{littlejohns_uk_2020}). These data points were collected starting from 2014 and preprocessed by the UKB. Utilizing a comprehensive 35-minute protocol, the UKB obtained many brain imaging modalities, including T1w and T2-FLAIR structural brain MRI images (\cite{smith_uk_2022}). We obtained all T1w and T2-FLAIR images available between 2014 and 2022 from 44,172 participants with neuroimaging data. 
 Raw T1w structural volumes were processed using a processing pipeline developed by UK Biobank researchers that consisted primarily of tools from FSL and Freesurfer. The pipeline generated additional images like segmentations between different types of matter and effectively reduced the non-brain tissue interference. Volumetric measures of gray matter and internal structures were generated alongside the processed images, providing valuable insights into the characteristics of gray matter and internal structures. Each T1w structural image underwent further processing with FreeSurfer (\cite{woolrich_bayesian_2009}), followed by a quality control check for inclusion into the data made available by UK Biobank researchers. Additionally, T2-FLAIR images were aligned to the corresponding T1 image, resulting in two additional product images. The UK Biobank saves volumetric measures of white matter lesions as additional data with both T1w and T2-FLAIR volumes. Images were reconstructed as DICOM images and converted to the NIfTI format using dcm2niix (\cite{li_first_2016}). All imaging volumes were defaced to preserve participant anonymity and broadcast to the MNI152 template space using FNIRT (\cite{woolrich_bayesian_2009}). Among these 44,172 participants, 43,369 participants have both T1w and T2-FLAIR images. To build 3D foundation models of neuroimages, we selected participants who had at least 100 slices in both T1w and T2-FLAIR volumes. This criteria resulted in 41,400 participants and 82,800 imaging volumes. A CONSORT diagram depicting the data used in this study can be found in Figure \ref{fig:consort}, and demographic data for participants is summarized in Figure \ref{fig:demo_data}. For detailed information, see the Appendix.

\begin{figure*}[!h]
\centering
\includegraphics[width=0.5\columnwidth]{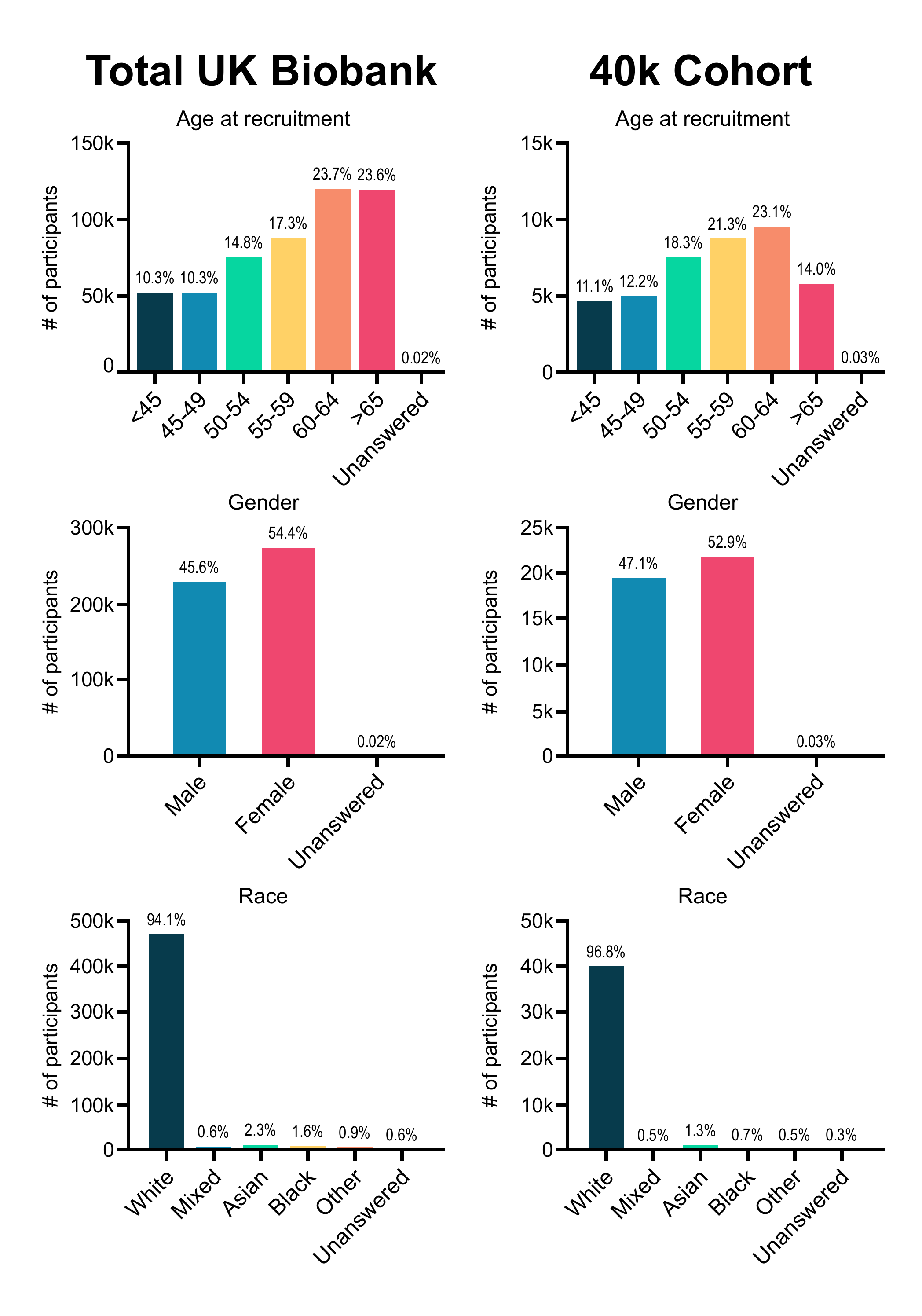}
\caption{Visual representation of demographic data from subjects in the UK Biobank in the study.}
\label{fig:demo_data}
\end{figure*}

\FloatBarrier

\subsubsection{BRaTS dataset}

 In Stage 2 and Stage 3, one of the datasets we used to perform self-supervised pretraining and finetuning on MRI images is from the training set of the BraTS 2021 Task 1 (Tumor Segmentation) challenge. This dataset consists of 1,251 subjects each with T1w, T1-contrast enhanced (T1-ce), T2-weighted (T2w), and T2-FLAIR images. We obtained all publicly available imaging volumes as part of the challenge. The BraTS challenge utilizes a standard preprocessing pipeline similar to the UKB dataset. First, images are converted from DICOM images to NIfTI using dcm2niix (\cite{li_first_2016}) and tools available from the Cancer Imaging Phenomics Toolkit (CaPTk) (\cite{davatzikos_cancer_2018}). Images are then co-registered to the SRI24 template and resampled to a uniform resolution of 1 mm\textsuperscript{3}. Finally, each modality is skull stripped, defaced, and converted to NIfTI (\cite{baid_rsna-asnr-miccai_2021}). Imaging volumes were then segmented into three tumor classes using the STAPLE algorithm (\cite{warfield_simultaneous_2004}) across previous BraTS winners and refined manually. These manual annotations were further verified by multiple board-certified neuro-radiologists, resulting in quality-controlled tumor segmentation labels across all four modalities in 3 classes: Gd-enhancing tumor (referred to as the whole tumor (WT), edematous tissue (ED), and necrotic tumor core (TC).

\subsubsection{ATLAS v2.0 Dataset}

Additionally, we perform self-supervised Stage 2 pretraining and fine-tuning on MRI images from the training set of the Anatomical Tracings of Lesions After Stroke (ATLAS) v2.0 Dataset \cite{liew_large_2022}. This dataset consists of 655 T1-ce MRIs aggregated from 44 research cohorts. Each MRI is from one subject, and time points range from $<$24 hours to $>$180 days after stroke onset. The standard labeling pipeline for the ATLAS dataset consists of (1) manual quality control to exclude significant motion artifacts, (2) manual lesion tracing in ITK-SNAP \cite{yushkevich_user-guided_2006} \cite{yushkevich_itk-snap_2017}, and (3) lesion mask review by two independent raters. The data then went through a similar prepossessing pipeline to BraTS. The MR images were intensity-normalized and registered to the MNI-152 template using the MINC toolkit (https://github.com/BIC-MNI/minc-toolkit). Finally, FreeSurfer's MRI deface functionality was used to deface the scans. Images were reviewed again in a final quality check at the end of the pipeline before being included in the dataset. Segmentations are evaluated on four metrics - Dice coefficient for the final segmentation (Dice), the difference between true total lesion volume and predicted total lesion volume (Volume Difference), the difference in the number of lesions between ground truth and prediction (Lesion Count), and Lesion-wise F1 Score. The Lesion-wise F1 Score is calculated by performing a 3D connected-component analysis to determine true positives, false positives, and false negatives. A true positive is any 3D connected component in the ground-truth image that overlaps with at least one voxel in the prediction image. Conversely, a false positive is any 3D connected component in the prediction image that does not overlap with the ground-truth image. A false negative is a connected component in the ground truth that lacks overlapping voxels in the prediction image \cite{liew_large_2022}. 

\subsection{Stage 1: Pretraining on the UKB}
 
 The initial pretraining stage involves the self-supervised learning of a transformer-based neural network model using a substantial unlabeled image dataset. For this purpose, the UKB dataset (\cite{littlejohns_uk_2020}) is utilized. From our 82,800 3D volumetric images used for pretraining, the input MRI modalities are randomly cropped into $96\times 96 \times 96$ sub-volumes and augmented with random inner cutout and rotation. These augmented images are then fed into the SwinUNETR encoder for processing.

The SwinUNETR architecture incorporates a Swin Transformer encoder 
that handles 3D input patches. This encoder operates with a patch size of $2\times 2\times 2$, a feature dimension of 8, and an embedding space of 48 dimensions. It consists of four stages, with a patch merging layer introduced between stages to reduce the feature size by half.

  Adopting the methodology from \cite{tang_self-supervised_2022}, the SwinUNETR encoder is pretrained through three distinct proxy tasks that serve as self-supervised fine-tuning mechanisms: masked volume inpainting, 3D image rotation, and contrastive coding. The primary objective of pretraining is to minimize the total loss function. 
  This work has developed three models to address varying complexities. These models include the foundational BrainSegFounder-Tiny with 62 million parameters, the intermediate BrainSegFounder-Small with 64 million parameters, and the advanced BrainSegFounder-Big with 69 million parameters. The primary differentiation among these models is the variation in the number of sliding window blocks within their third stage. Table~\ref{tab:encoder} shows BrainSegFounder's sliding-window encoder backbone's parameters, number of SSL heads, and number of sliding window blocks. 

 For the pretraining process, 64 NVIDIA DGX A100 GPUs, distributed across 8 DGX-2 nodes, are deployed at the University of Florida's HiPerGator-AI supercomputer. Data parallelism is implemented to optimize the efficiency of model training. Both training and validation losses are monitored to track progress. The AdamW optimizer is employed using a warm-up cosine scheduler set for 500 iterations. The training employs a batch size of 2 per GPU, using $96\times 96\times 96$ patches. The initial learning rate is established at $6\times 10^{-6}$, coupled with a momentum of 0.9 and a decay of 0.1 over 15,000 iterations. These parameters are summarized in Table \ref{tab:hyperparameter-settings}.

\begin{table*}[!t]
    \small
    \setlength{\tabcolsep}{4pt}
    \centering
    \resizebox{\textwidth}{!}{
    \begin{tabular}{|l|l|l|l|l|l|l|l|}
        \hline
        \multicolumn{2}{|l}{} & \multicolumn{2}{|l|}{\textbf{BrainSegFounder-Tiny (62M)}} & \multicolumn{2}{l|}{\textbf{BrainSegFounder-Small (64M)}} & \multicolumn{2}{|l|}{\textbf{BrainSegFounder-Big (69M)}} \\
        \hline
        \multicolumn{2}{|l|}{\# of encoder parameters} & \multicolumn{2}{l|}{19,097,191} & \multicolumn{2}{l|}{20,982,103} & \multicolumn{2}{l|}{26,636,839} \\
        \hline
        Encoder layer level & Output size

        & \# of SSL Heads  & \# Swin Blocks & \# of SSL Heads & \# Swin Blocks & \# of SSL Heads  & \# Swin Blocks   \\
        \hline
        Level 1 & 48x(48x48x48) & 3 & 2 & 3 & 2 & 3 & 2 \\
        \hline
        Level 2 & 96x(24x24x24) & 6 & 2 & 6 & 2 & 6 & 2 \\
        \hline
        Level 3 & 192x(12x12x12) & 12 & 2 & 12 & 6 & 12 & 18 \\
        \hline
        Level 4 & 384x(6x6x6) & 24 & 2 & 24 & 2 & 24 & 2 \\
        \hline
        \multicolumn{2}{|l|}{Feature size} & \multicolumn{2}{l|}{48} & \multicolumn{2}{l|}{48} & \multicolumn{2}{l|}{48} \\
        \hline
        \multicolumn{2}{|l|}{Bottleneck dimension} & \multicolumn{2}{l|}{768} & \multicolumn{2}{l|}{768} & \multicolumn{2}{l|}{768} \\
        \hline

    \end{tabular}}
    \caption{Pretraining encoder settings.}
    \label{tab:encoder}
\end{table*}

\begin{figure*}[!t]
\centering
\includegraphics[width=\textwidth]{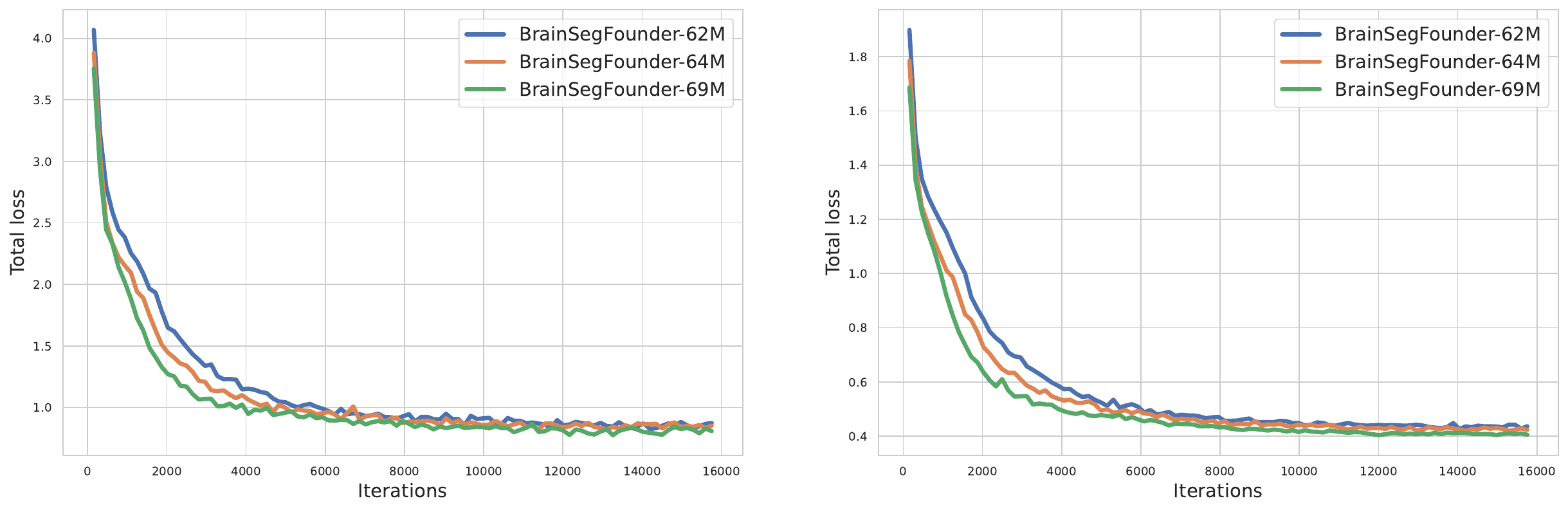}
\caption{Training (left) and validation (right) loss of Stage 1-pretraining three different scale of BrainSegFounder models on UKB.}
\label{fig:stage_1_pretrain_curves}
\end{figure*}

\subsection{Training on BraTS}
\subsubsection{Stage 2: Pretraining on BraTS}

 The pretrained models based on the UK Biobank (UKB) dataset underwent further pretraining through transfer learning on the Brain Tumor Segmentation (BraTS) dataset. 1,251 subjects were employed for a 5-fold cross-validation process. To ensure consistent performance evaluation, the data splits for these 5 folds were kept identical to those used in the baseline SwinUNETR model. During training, four of the folds were utilized for training purposes, and the remaining folds served for validation.

 Given that the BraTS dataset comprises four modalities, but only two (specifically, T1w and T2-FLAIR) were available for pretraining in the initial stage, the first layer of the pretrained network on UKB was modified. This modification involved expanding the number of input channels by adding two new channels, whose weights were randomly initialized using the Kaiming initialization method (\cite{he_delving_2015}).

 Hyperparameter settings for Stage-2 Pretraining can be found in Table \ref{tab:hyperparameter-settings}. For pretraining on BraTS, two NVIDIA A100 GPUs, each with 32 GB of memory, were utilized. Depending on the model size, the BrainSegFounder models require between 48 to 72 hours for training. The batch size and learning rate were uniformly set at 2 and $1\times10^{-4}$ for all models during this pretraining phase.

\subsubsection{Stage 3: Fine-tuning on BraTS} 
In the final fine-tuning stage we attach the pretrained encoder from the previous stage to a UNet decoder. This model is then finetuned directly on the BraTS dataset. We used the same hyper-parameter settings as those used in the Stage 2 pretraining phase on BraTS (in Table \ref{tab:hyperparameter-settings}): The batch size remained at 2, mirroring the encoder-only stage, and the learning rate remained at $1\times10^{-4}$. The number of steps for this phase was set to 50,000, with the input data having 4 channels, which indicates the typical inclusion of multi-modal MRI scans in the BraTS dataset.

\subsubsection{Few-shot Learning on BraTS}
To investigate our model's performance using limited training data, we conducted a systematic comparison between BrainSegFounder and the baseline model, SwinUNETR, utilizing a descending percentage training approach in the context of the BraTS challenge. Using both our BrainSegFounder pretrained model and SwinUNETR, we finetuned on 40\% of the BraTS training dataset, with subsequent incremental reductions in data availability, decreasing to a final 5\% of the original dataset. Due to the potential high-variability on the input dataset at such small percentages of input data, we trained on 5 different randomly sampled subsets of the input training data and calculated the average performance across these subsets. This method aimed to explore the impact of training data scarcity on model performance and adaptability. Performance evaluations were carried out on the BraTS test set after each training step and evaluated with the Dice coefficient to assess segmentation accuracy. 

\begin{table*}[!t]
    \caption{Hardware and training parameters.}
    \centering
    \small
    \setlength{\tabcolsep}{4pt} 
    \resizebox{\textwidth}{!}{\begin{tabular}{|l|l|l|l|l|l|l|l|l|}
    \hline
    & \textbf{Stage} & \textbf{Data} &\textbf{ No. Subjects} & \textbf{GPU}  & \textbf{Batch size }& \textbf{Learning rate} & \textbf{No. steps}  & \textbf{No. input channel} \\
    \hline
    Encoder only  & Stage 1 Pretraining & UKB  & 43369 & 64 $\times$ A100 & 128 & $1 \times 10^{-6}$ & $2 \times 10^5$ & 2 \\
    \hline
    Encoder + Decoder & Stage 2 Pretraining \& Stage 3Fine-tuning & BRaTS & 1251 & 2 $\times$ A100  & 2 & $1 \times 10^{-4}$ & 50000 & 4 \\
    \hline
    Encoder + Decoder & Stage 2 Pretraining \& Stage 3 Fine-tuning & ATLAS v2.0 & 655 & 4 $\times$ A100  & 4 & $3 \times 10^{-3 }$& 600 & 1 \\
\hline

\end{tabular}}
    \label{tab:hyperparameter-settings}
\end{table*}

\subsubsection{Modality Restriction and Flexibility in Training and Inference}
The proposed method is designed to be adaptable across various data modalities in downstream tasks. In the case that fewer modalities are available in the downstream task, Stage 1 pretrained model using both T1- and T2-weighted MRI can be adapted and fine-tuned on fewer modalities (e.g., T1- or T2- weighted MRI alone) during Stage 2 pretraining, Stage 3 supervised training, and the inference without requiring any modifications to the network structure. This is achieved by simply configuring the two input channels to process the same type of data (either T1- or T2-weighted).

In the case that more modalities are available in the downstream task, our model can also accommodate this by increasing the number of input channels. The pre-trained weights are then loaded into the corresponding layers of the network.

To investigate the efficacy of this method, we performed ablation testing on BraTS by restricting the modalities available to the model. The Stage 1 model was given only either T1w or T2w images rather than utilizing both modalities available from the UKBiobank. These models, pretrained on only one modality, were then trained with our Stage 2 pretraining pipeline on the BraTS data with only the modality trained on in Stage 1 instead of the 4 modalities available from BraTS. Finally, the model was finetuned on BraTS with the same single modality. Hyperparameters and number of GPUs were kept the same as in our earlier pretraining steps on BraTS as summarized in Table \ref{tab:hyperparameter-settings}.

\subsection{Training on ATLAS v2.0}
\subsubsection{Stage 2: Pretraining on ATLAS v2.0}

Similarly, the pretrained models based on the UK Biobank (UKB) underwent further self-supervised pretraining through transfer learning on the ATLAS v2.0 dataset. In this stage, a total of 665 MR images were included in the training set, intentionally avoiding cross-validation to align our methodology with that employed by the submissions in the challenge leaderboard. This approach allows for direct performance comparisons under similar training conditions, providing a robust test of our models against established benchmarks.

Since the ATLAS dataset has only one modality, T1-ce, the first layers of the Stage-1 pretrained model were modified by dropping the channel corresponding to the T2w modalitiy present in the UKB. For pretraining, four NVIDIA A100 GPUs, each with 32 GB of memory, were utilized. The stage 2 model took 35 hours to train, with a batch size and learning rate set to 4 and $5\times10^{-3}$, respectively. 

\subsubsection{Stage 3: Fine-tuning on ATLAS v2.0}
Upon completion of pretraining, the model was fine-tuned further on the ATLAS v2.0 dataset to adapt to the specific challenges of lesion detection in stroke patients. The fine-tuning employed a cosine-annealing learning rate scheduler, starting with an initial learning rate of $1\times10^{-4}$. Batch size was set to 4, and the model was trained for a total of 600 epochs.

Training was conducted using 2 NVIDIA A100 GPUs with 32GB of RAM accessible to each GPU. We applied data augmentation techniques of random cropping, rotation, and to improve model robustness against variations in real-world data. The loss function used was Dice-loss, suited for addressing imbalance between classes. Dropout at a rate of 10\% was applied to prevent overfitting.

Model performance was periodically assessed using a held-out set of 100 images from the training dataset, ensuring that the model's improvements were generalizable and aiding with tuning the hyperparameters effectively and tracking training progress without the use of a separate validation dataset.

\section{Results}
\subsection{Pretraining}
 The pretraining of our BrainSegFounder models, which varied in size based on the number of parameters, took between 3 to 6 days. This process utilized a computational setup ranging from 8 to 64 NVIDIA A100 GPUs, each with 80 GB capacity. Figure \ref{fig:stage_1_pretrain_curves} illustrates the validation loss during the pretraining phase across different BrainSegFounder model sizes. 
 
\subsection{Evaluation on BraTS Challenge Dataset}
\subsubsection{Comparison to State-of-the-Art Methods}
 Table \ref{tab:GB_soa_comparison} summarizes our BrainSegFounder's best performing model against published results on our validation splits from other state-of-the-art models on the BraTS challenge: the baseline SwinUNETR model \cite{hatamizadeh_swin_2022}, nnU-Net \cite{isensee_nnu-net_2021}, TransBTS \cite{wang_transbts_2021}, SegResNet \cite{myronenko_3d_2018_segresnet}, and MONAI's Model-Zoo \cite{project-monaimodel-zoo_2024}. SegResNet and nnU-Net are both winning methodologies in previous BraTS challenges. TransBTS is a vision-transformer based approach tailored for brain tumor segmentation. MONAI's Model-Zoo is a bundle of medical imaging models capable of performing a wide variety of tasks, including BraTS segmentation. In addition, we include comparison to the corresponding single-stage pretrained model that was pretrained only on the UKB. Table \ref{tab:GB_comparison} in the appendix present a comparative analysis of all trained BrainSegFounder models (of varying sizes) against the current leading model in this field broken down by model size.

\begin{table*}[!t]
\caption{A comparison of BrainSegFounder (BSF) models' performance in terms of average Dice coefficient on the BraTS challenge. BSF-S indicates our best performing BrainSegFounder model (small, 64M parameters). BrainSegFounder models were pretrained with SSL on T1- and T2-weighted MRI 3D volumes and finetuned with supervised learning using all four modalities present in BraTS. BSF-1-S indicates this model with only the Stage 1 (SSL) pertaining on UKB and without the Stage 2 pretraining step. SwinU models are models using the SwinUNETR architecture trained on BraTS via supervised learning. SwinU-MRI is the model trained directly using supervised learning on BraTS published on GitHub (\url{https://github.com/Project-MONAI/research-contributions/tree/main/SwinUNETR/BRATS21}), SwinU-Res is pretrained with SSL on only T1w and T2w and finetuned on BraTS, and SwinU-CT pretrained using CT data and finetuned with supervised learning on BraTS. nnU-Net and SegResNet are former BraTS challenge winners trained using supervised learning on our folds. TransBTS is a vision-transformer based segmentation algorithm optimized for brain segmentation. Model-Zoo is a bundle of models published by MONAI that can perform BraTS segmentation out of the box using their "Brats mri segmentation" [sic] model  found at \url{https://monai.io/model-zoo.html}}
\label{tab:GB_soa_comparison}
\small
\centering

\resizebox{\textwidth}{!}{\begin{tabular}{| l | l | l | l | l | l | l | l | l | l |}
\hline
 & \textbf{BSF-S (64M)} & \textbf{BSF-1-S (64M)} & \textbf{SwinU-MRI}  & \textbf{SwinU-Res} &  \textbf{SwinU-CT} & \textbf{nnU-NeT} & \textbf{SegResNet} & \textbf{TransBTS} & \textbf{Model-Zoo} \\
\hline
Fold 1 & \textbf{0.9032} & 0.8994 & 0.8854 & 0.895 & 0.894 & 0.896 & 0.899 & 0.883 & 0.857 \\
\hline
Fold 2 & \textbf{0.9182} & 0.9055 & 0.9059 & 0.899 & 0.902 & 0.917 & 0.916 & 0.902 & 0.879 \\
\hline
Fold 3 & 0.9121 & \textbf{0.9125} & 0.8981 & 0.894 & 0.898 & 0.910 & 0.909 & 0.889 & 0.820 \\
\hline
Fold 4 & 0.9100 & \textbf{0.9133} & 0.8924 & 0.890 & 0.893 & 0.909 & 0.908 &  0.893 &  0.889 \\
\hline
Fold 5 & \textbf{0.9139} & 0.9114 & 0.9035 & 0.903 & 0.902 & 0.909 & 0.906 & 0.892 & 0.893 \\
\hline
Average & \textbf{0.9115} & 0.9110 & 0.8971 & 0.896 & 0.898 & 0.908 & 0.907 & 0.891 & 0.868 \\
\hline
\end{tabular}}
\label{tab:bsf-comparison}
\end{table*}

These results show that pretraining on a large scale of healthy brain MRI data from UKB can significantly improve performance. Other observations can be made as below:

 \begin{itemize}
    \item Across all folds, the BrainSegFounder-Small (BSF-S) framework consistently outperformed the SwinUNETR model. This indicates that the additional training steps taken within the BrainSegFounder framework play a significant role in enhancing its effectiveness in brain tumor segmentation tasks.
    \item The Small (64M parameters) version of BrainSegFounder achieved higher Dice coefficients on average than the 62M and 69M versions. We believe this indicates that there is an optimal range of model complexity that maximizes performance given certain training data size. Simply increasing the number of parameters does not necessarily lead to better results if the training data does not scale up. 
    \item The one-stage Tiny (62M parameters) model performed comparably to the two-stage BrainSegFounder-Tiny (62M parameters) model, which is notable, implying it did not benefit considerably from the second stage pretraining on the BraTS. This might imply that the UKB dataset alone provides enough variability for effective training. Further study should be made to verify whether the benefit of pretraining on the target datasets can be found using large-scale networks.

\end{itemize}
\subsubsection{Few-shot learning}
\begin{figure}
    \centering
    \includegraphics[width=\columnwidth]{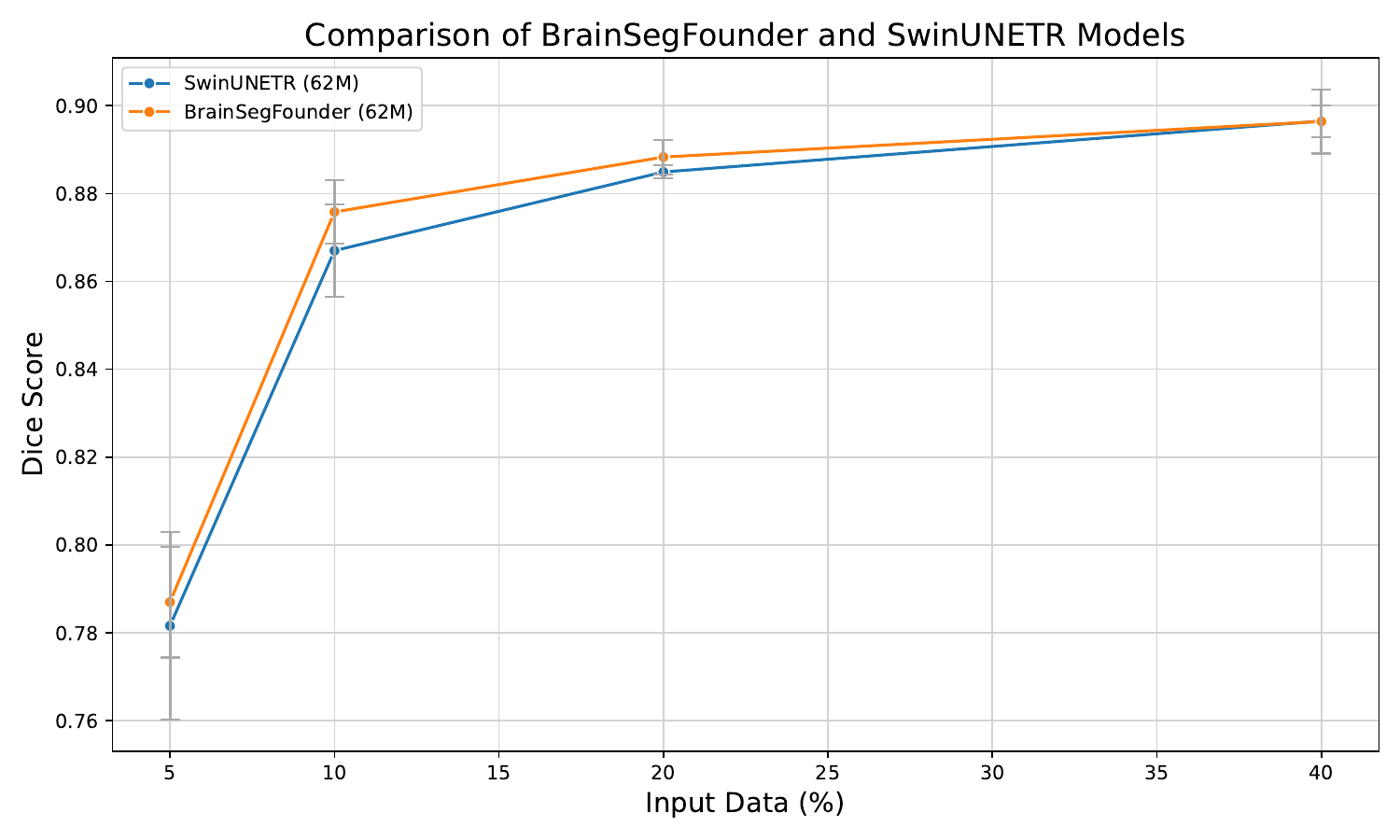}
    \caption{Dice coefficients for baseline (SwinUNETR) and our model across different levels of training data availability. All models were trained 5 times to account for variability in the input data randomly selected. Error bars represent $\pm$ one standard deviation.}
    \label{fig:fewshot}
\end{figure}

Our experimental results demonstrate the performance capabilities of BrainSegFounder relative to the baseline model, SwinUNETR, under constrained training data conditions. As depicted in Figure \ref{fig:fewshot}, BrainSegFounder consistently matched the performance of SwinUNETR across higher levels of available training data and outperformed SwinUNETR when training data was constrained to lower input percentages. As the percentages of training data approached  40\% of the input data, both models achieved nearly equivalent accuracy. However, as the amount of training data decreased, BrainSegFounder exhibited superior robustness and adaptability. Notably, at all data availability levels, BrainSegFounder maintained higher mean segmentation accuracy. The results presented in Figure \ref{fig:fewshot} for these input levels are an average of 5 independent subsets of input data to account for variability in small datasets. Supplemental Table \ref{tab:brainseg_comparison} contains the results for each of these subsets.

Overall, the BrainSegFounder (64M) model provides the best balance between complexity and performance, as evidenced by its leading average Dice coefficient. These results demonstrate the potential benefits of pretraining on the data from a large number of health subjects. 

\subsubsection{Modality Restriction}

Our model can effectively adapt with fewer or single modality data in the downstream task. Table \ref{tab:modality-restriction} shows the five-fold cross-validation and average Dice scores comparing SSL-pretrained BrainSegFounder and fully supervised SwinUNETR, both using T1-weighted MRI only for all training and inference stages. Our multi-stage pretraining demonstrated a DICE score improvement of 0.04 (6\%), indicating the substantial benefit of multi-stage pretraining when the downstream task has limited data modality.

\begin{table*}[h]
    \centering
    \begin{tabular}{|l|l|l|l|}
        \hline
         & \textbf{BrainSegFounder (T1w only)} & \textbf{SwinUNETR (T1w only)} \\ \hline
         Fold 1 & \textbf{0.718} & 0.700 \\ \hline
         Fold 2 & \textbf{0.721} & 0.672 \\ \hline
         Fold 3 & \textbf{0.707} & 0.657 \\ \hline
         Fold 4 & \textbf{0.731} & 0.686 \\ \hline
         Fold 5 & \textbf{0.725} & 0.676 \\ \hline
         Average & \textbf{0.721} & 0.678 \\ \hline
    \end{tabular}
    \caption{Performance comparison of modality restricted models on the BraTS dataset. SwinUNETR is fully supervised learning on T1-weighted MRI without pretraining, while BrainSegFounder uses our multi-stage pretraining on UKB and BraTS T1-weighted MRI and is then finetuned on BraTS T1-weighted MRI.}
    \label{tab:modality-restriction}
\end{table*}

\subsection{Evaluation on ATLAS Challenge Dataset}
Our model's performance on the ATLAS v2.0 dataset was compared against the top-performing models listed on the challenge leaderboard. The results, as summarized in Table \ref{tab:atlas_comparison}, demonstrate that our model achieved a Dice score of 0.712, a lesion-wise F1-score of 0.711, simple lesion count of 3.421, and volume difference of 8993.85. These scores would place our model within the top 3 models in the training set leaderboard. Worth noting again is that our training protocol did not include 5-fold cross-validation due to the lack of predetermined folds for which ATLAS can be evaluated in the training set. Top-performing models on the leaderboard\footnote{\url{https://atlas.grand-challenge.org/evaluation/lesion-segmentation-hidden-test-set/leaderboard/}} (CTRL, HeRN, and POBOTRI) were not available for independent validation. Instead, our approach focused on maximizing the comparability with the leaderboard conditions by adhering closely to their reported training setups.

\begin{table*}[h]
    \centering
    \begin{tabular}{|l|l|l|l|l|l|}
    \hline
    \textbf{Metric} & \textbf{BrainSegFounder} & \textbf{SwinUNETR} & \textbf{CTRL(*†)} & \textbf{HeRN(*‡)} & \textbf{POBOTRI(*)}\\
    \hline
    \textbf{Dice} (↑) & 0.712 & 0.703 & 0.663 & 0.718 & 0.663 \\ \hline
    \textbf{Lesion-wise F1 Score} (↑) & 0.711 &  0.703 & 0.556 & 0.724 & 0.559 \\ \hline
    \textbf{Simple Lesion Count} (↓) & 3.421 & 3.677 & 4.657 & 2.750 & 4.500 \\ \hline
    \textbf{Volume Difference} (↓) & 8993.85 & 9165.18 & 8804.91 & 6162.00 & 9535.23 \\ \hline
    \end{tabular}
    \caption{Performance comparison of segmentation models on the ATLAS v2.0 dataset. All metrics from the challenge (Dice coefficient, Lesion-wise F1 Score, Simple Lesion Count, and Volume Difference) are included for each model for each model. Scores for models marked with an asterisk (*) are sourced directly from the official challenge leaderboard and pertain to their performance on the training set. All of these models utilize ensemble learning methods. CTRL (denoted with †) is the official challenge winner, while HeRN (denoted with ‡) leads on the training set. Top-performing models on the leaderboard (CTRL, HeRN, and POBOTRI) were unavailable for independent validation. }
    \label{tab:atlas_comparison}
\end{table*}

\section{Discussion}

The findings from our work with BrainSegFounder, especially the "Small" model comprising 64 million parameters, signify a noteworthy progression in 3D foundation models for neuroimage segmentation. The framework's novel two-stage pretraining strategy—initially utilizing a broad dataset of multi-modal neuroimages from the generally healthy population found in the UK Biobank, followed by training on diseased brain MRI volumes from the BraTS dataset—has demonstrated substantial efficacy. The first stage of pretraining enables the framework to capture the latent representation of normal brain anatomy, a vital aspect for the precise identification of anomalies, like those encountered in brain tumor segmentation tasks. The second stage of pretraining learns the spatial distribution and texture representations of lesions presented in different brain disorders. 

Table~\ref{tab:bsf-comparison} demonstrates superior performance of BrainSegFounder compared to state-of-the-art approaches including SwinUNETR, nn-UNET, SegResNet, TransBTS, and the brain segmentation foundation model published in Model-Zoo. Each of these models represents a significant advancement in brain segmentation techniques. SwinUNETR is the most direct comparison - its architecture is identical to ours, and outperforming SwinUNETR on these tasks indicates that our multi-stage, self-supervised pretraining method is effective in improving segmentation performance. nn-UNeT and SegResNet both were former BraTS challenge winners. As such, they are empirically validated models that excel in the BraTS challenge. By outperforming these models, BrainSegFounder demonstrates its capability to perform segmentation tasks with exceptionally high accuracy and precision. TransBTS utilizes both convolutional and transformer-based architectures to Swinoth local and global context; our superior results compared to TransBTS demonstrate that our model's performance increase is not merely due to the inclusion of a transformer based architecture and further validate our pipeline. The brain segmentation foundation model from Model-Zoo serves as a comprehensive pre-trained model designed specifically for brain imaging tasks. By demonstrating increased results compared to this novel foundation model, we show that our methodology could streamline the creation of more effective medical foundation models for segmentation.

One of the key findings is the superior performance of the BrainSegFounder-Small (64M) model over other BrainSegFounder variants. Based on our limited explored range of parameters, our model performs best with an intermediate number of parameters. This suggests that an optimal balance of model complexity and training data is crucial. It is also indicative of the importance of large-scale datasets in training 3D vision foundation models for medical imaging, as even the one-stage pretraining model showed significant effectiveness.  However, there is still a possibility to see higher performance using a higher number of parameters and large-scale diverse training data that we did not explore in this work.

Further, the comparable performance of the one-stage 62M model with the two-stage approach indicates that extensive pretraining on a large and diverse dataset like UKB might be sufficient for effective model training, reducing the need for additional pretraining on targeted datasets. This insight could streamline future 3D foundation model development for medical imaging, especially in scenarios where specific pathological datasets are limited or hard to acquire.

Our results from our study limiting training data in few-shot learning indicate that BrainSegFounder's training methods potentially offer better generalization from limited data, a crucial factor for practical applications in medical imaging where annotated data can be scarce. Though only a slight improvement, the BrainSegFounder consistently outperforms the baseline model at lower levels of input data (see Fig \ref{fig:fewshot} and Supplemental Table \ref{tab:brainseg_comparison}). Even with incredibly limited data, our Stage 2 self-supervised pretraining serves as a meaningful inclusion in the training pipeline. These findings suggest that the enhancements integrated into BrainSegFounder are effective in optimizing performance under varying data constraints, thereby affirming its suitability for real-world deployment in medical imaging contexts.

In our modality restriction experiment, our model sees a significant reduction in quality when training with fewer modalities. This drop indicates that the multiple modalities present in BraTS contain important information not present in just T1-weighted MRI images about tumor segmentation. However, BrainSegFounder's better performance under these more challenging scenarios with limited modality input when compared to the base SwinUNETR model validates the feasibility of our extensible approach to handling varying numbers of modalities. When fewer modalities are present, our training scheme still provides valuable information and performance improvements by keeping only the layers trained on the modality present. Similarly, our results using all four modalities present in BraTS suggest our method effectively utilizes information given in the pretraining step when presented with additional modalities. Therefore, we conclude that the pretraining steps have a positive effect even when the model is provided with more or less information than is present in the original pretraining stage. Moreover, our results on ATLAS (discussed below) further support our method of handling multiple modalities. 

The performance of BrainSegFounder on the ATLAS dataset indicates that its training scheme is generalizable and effective at more than just tumor segmentation, a trait desirable for foundation models. While methods specifically adapted to optimizing results on this dataset do outperform ours, we still maintain third place in the leaderboard. Remarkably, our results were achieved without the use of ensemble learning techniques, which are commonly employed to boost performance by leveraging the strengths of multiple models. The fact that our single-model approach is competitive with ensemble models underscores the robustness and efficiency of our model in managing the intricacies of medical image analysis. We believe that the methodology used for BrainSegFounder can be refined and extended to move towards a Medical Foundation Model for neuroimages. 

In addition, BrainSegFounder's training scheme and model provide a clear advantage over SAM and MedSAM, two powerful existing segmentation foundation models. (1) While SAM is restricted to 2D RGB images, BrainSegFounder is designed to handle 3D medical images with any number of channels as input, providing greater versatility in medical imaging applications. (2) MedSAM requires bounding-box input prompts and its 3D functionality is limited to manually uploading each image to a plugin for prompting and slice-by-slice annotation. Both methods require manual input. In contrast, our model eliminates the need for such manual interventions once trained, streamlining the segmentation process. (3) Although SAM is capable of automated segmentation without input, it lacks the ability to specify a fixed number of classes and instead generates an arbitrary number of classes; this property leads sub-optimally on medical images with specific segmentation tasks (e.g., lesion detection), and cannot be used without additional human input. (4) Neither SAM nor MedSAM efficiently process multimodal data as they generate predictions from a single scan, whereas BrainSegFounder is designed to integrate multiple scans from the same individual. 

However, it's important to note that while BrainSegFounder shows promise in brain tumor segmentation and brain region segmentation, its application in other neuroimaging tasks remains to be explored. One such task is brain tissue segmentation - a common task in automated analysis. Future research should investigate its adaptability to other neurological conditions, its performance in different clinical environments, and its usefulness in additional common analysis tasks.

In conclusion, BrainSegFounder is a significant step forward 3D foundation models for medical image segmentation and analysis, particularly for multi-modal neuroimaging. Its development underscores the potential of AI and foundation models in enhancing diagnostic accuracy and efficiency, paving the way for more advanced, adaptable, and robust AI tools in healthcare.

\section{Acknowledgments}

This work was partially supported by the National Science Foundation (1908299, 2318984, 2123809) and the National Institute of Aging of the National Institutes of Health (RF1AG071469). This research has been conducted using the UK Biobank Resource under application number 48388. We extend our appreciation to the contributors of the UK Biobank, BraTS, and ALTAS datasets, whose extensive data collection efforts have been fundamental to the success of this research. We gratefully acknowledge NVIDIA AI Technology (NVAITC) 's support for this research project, especially on GPU parallelization technology. Special thanks to the University of Florida's HiPerGator-AI supercomputer team for providing the computational resources essential in training the BrainSegFounder models. We appreciate the support from Ying Zhang (UFIT Research Computing) and Kaleb Smith (NVAITC) in providing the necessary computing resources and technology support. 

\printbibliography

\section{Appendix}
\subsection{UK Biobank Data}
Table \ref{tab:UKB_demo} presents a comprehensive summary of the participants used from the UK Biobank.
\begin{table*}[!t]
    \centering
    \begin{tabular}{| l | l | l |}
        \hline
        & \textbf{Entire UK Biobank} & \textbf{40K Cohort (\%)} \\
        \hline
        \hline
        \multicolumn{1}{l}{ } & \multicolumn{1}{l}{Age at Recruitment} & \multicolumn{1}{l}{ }\\
        \hline
        $\leq$ 45 & 51,763 (10.3\%) & 4,601 (11.1\%) \\
        \hline
        45-49 & 51,866 (10.3\%) & 5,031 (12.2\%) \\
        \hline
        50-54 & 74,387 (14.8\%) & 7,563 (18.3\%) \\
        \hline
        55-59 & 86,899 (17.3\%) & 8,819 (21.3\%) \\
        \hline
        60-64 & 118,959 (23.7\%) & 9,579 (23.1\%) \\
        \hline
        $\geq$ 65 & 118,435 (23.6\%)  & 5,796 (14.0\%) \\
        \hline
        Unanswered & 101 (0.02\%) & 11 (0.03\%) \\
        \hline
        \multicolumn{1}{l}{ } & \multicolumn{1}{l}{Gender} & \multicolumn{1}{l}{ }\\
        \hline
        Male & 229,051 (45.6\%) & 19,497 (47.1\%) \\
        \hline
        Female & 273,258 (54.4\%) &  21,891 (52.9\%) \\
        \hline
        Unanswered & 101 (0.02\%) & 12 (0.03\%) \\
        \hline
        \multicolumn{1}{l}{ } & \multicolumn{1}{l}{Race} & \multicolumn{1}{l}{ }\\
        \hline
        White & 472521 (94.1\%) & 40057 (96.8\%) \\
        \hline
        Mixed & 2953 (0.6\%) & 190 (0.5\%) \\
        \hline
        Asian & 11447 (2.3\%) & 541 (1.3\%) \\
        \hline
        Black & 8055 (1.6\%) & 268 (0.7\%) \\
        \hline
        Other & 4555 (0.9\%) & 223 (0.5\%) \\
        \hline
        Unanswered & 2,778 (0.6\%) & 121 (0.3\%) \\
        \hline
        \multicolumn{1}{l}{ } & \multicolumn{1}{l}{Data Information} & \multicolumn{1}{l}{ }\\
        \hline
        \# Samples & 502,309 & 41,400 \\
        \hline
        \# Brain Tumors & 1,210 & 42\\
        \hline
    \end{tabular}
    \caption{UKB Data Demographic information.}
    \label{tab:UKB_demo}
\end{table*}

\subsection{Fold-wise comparison of all models.}
Table \ref{tab:GB_comparison} provides fold-wise comparison of our BrainSegFounder models across all tested parameters.
\begin{table*}[!t]  
    \small
    \centering
    \resizebox{\textwidth}{!}{
    \begin{tabular}{| l | l | l | l | l | l | l | l |}
        \hline
         & \textbf{SwinUNETR (62M)} & \textbf{BSF-T (62M) }&  \textbf{BSF-S}  (64M) &  \textbf{BSF-B}  (69M) &\textbf{ One-Stage (62M)} & \textbf{One-Stage (64M)} & \textbf{One-Stage (69M)} \\
        \hline
        Fold 1& 0.8854 & 0.9027 & \textbf{0.9032} & 0.9014 & 0.9019 & 0.8994 & 0.8999 \\
        \hline
        Fold 2 & 0.9059 & 0.9181 & 0.9182 & 0.9164 & \textbf{0.9188} & 0.9186 & 0.9055 \\
        \hline
        Fold 3& 0.8981 & 0.9102 & \textbf{0.9121} & 0.9097 & 0.9119 & 0.9125 & 0.9002 \\
        \hline
        Fold 4 & 0.8924 & 0.9103 & 0.9100 & 0.9070 & 0.9107 &  \textbf{0.9133} &  0.9109 \\
        \hline
        Fold 5& 0.9035 & 0.9139 & \textbf{0.9141} & 0.9101 & 0.9132 & 0.9114 & 0.9103 \\
        \hline
        Average & 0.8971 & 0.9110 & \textbf{0.9115} & 0.9089 & 0.9112 & 0.9110 & 0.9054 \\
        \hline
    \end{tabular}}
    \caption{Comparison of BrainSegFounder models through 5-fold cross-validation with metric Dice coefficient on BraTS. SwinUNETR is the winning solution on BraTS challenge 2021, which is performed with fully supervised learning without UKB pretraining.  BrainSegFounder is the proposed method, which is conducted with the two-stage pretraining and then finetuning on the target dataset. The one-stage means that pretraining on UKB is performed but not on the BraTS. Note: The performance results for SwinUNETR were published on the official GitHub, utilizing hyper-parameter settings similar to those in our finetuning stage but without implementing the ensembling approach that was described in the published work.}
    \label{tab:GB_comparison}

\end{table*}

Table \ref{tab:brainseg_comparison} presents a comparison of few-shot learning Dice scores on the testing set at varying levels of input training data.
\begin{table*}[!t]
\centering
\begin{tabular}{|c|c|c|c|c|c|c|c|c|}
\hline
\multirow{2}{*}{Repeat} & \multicolumn{2}{c|}{5\%} & \multicolumn{2}{c|}{10\%} & \multicolumn{2}{c|}{20\%} & \multicolumn{2}{c|}{40\%} \\ \cline{2-9}
 & BSF & SwinU & BSF & SwinU & BSF & SwinU & BSF & SwinU \\ \hline 
1 & \textbf{0.7810} & 0.7437 & \textbf{0.8771} & 0.8594 & \textbf{0.8893} & 0.8837 & \textbf{0.8949} & 0.8885\\ \hline
2 & \textbf{0.7912} & 0.7899 & \textbf{0.8764} & 0.8553 & 0.8814 & \textbf{0.8834} & \textbf{0.8981} & \textbf{0.8981} \\ \hline
3 & 0.7797 & \textbf{0.7886} & 0.8683 & \textbf{0.8812} & \textbf{0.8901} & 0.8869 & 0.8906 & \textbf{0.8956} \\ \hline
4 & \textbf{0.8071} & 0.7966 & 0.8705 & \textbf{0.8735} & \textbf{0.8911} & 0.8844 & \textbf{0.9048} & 0.8938 \\ \hline
5 & 0.7758 & \textbf{0.7893} & \textbf{0.8869} & 0.8656 & \textbf{0.8895} & 0.8860 & 0.8857 & \textbf{0.8938}\\ \hline 
Average & \textbf{0.7870} & 0.7816 & \textbf{0.8758} & 0.8670 & \textbf{0.8883} & 0.8849 & \textbf{0.8948} & 0.8937 \\ \hline
\end{tabular}
\caption{Comparison of BrainSegFounder (BSF) and SwinUNETR (SwinU) Baseline models trained on 5 repeats of varying percentages of the input data. Data was randomly sampled from the BraTS training dataset, and models were evaluated on the testing dataset.}
\label{tab:brainseg_comparison}
\end{table*}

\end{document}